\documentclass[twocolumn]{aastex631}

\usepackage{makecell}
\usepackage{comment}
\usepackage{natbib}
\usepackage{multirow}
\usepackage{booktabs}
\usepackage{epsfig}
\usepackage{ulem}
\usepackage{rotating}
\usepackage{graphicx}
\usepackage{url}
\usepackage{xcolor}
\usepackage{amsmath}
\usepackage{color}
\usepackage{savesym}
\savesymbol{tablenum}
\usepackage{siunitx}
\usepackage[normalem]{ulem}
\usepackage{appendix}
\usepackage{enumerate}
\usepackage{gensymb}
\usepackage{longtable}
\usepackage{afterpage}
\usepackage{threeparttable}
\usepackage{tabularx}
\usepackage[bottom]{footmisc}

\DeclareSIUnit\parsec{pc}
\DeclareSIUnit\h{\textit{h}}

\begin{document}

    \title{The Hubble Tension in our own Backyard: DESI and the Nearness of the Coma Cluster}
    
    \author[0000-0002-4934-5849]{Daniel Scolnic}
    \affiliation{Department of Physics, Duke University, Durham, NC 27708, USA}
    
    \author[0000-0002-6124-1196]{Adam G.~Riess}
    \affiliation{Space Telescope Science Institute, Baltimore, MD 21218, USA}
    \affiliation{Department of Physics and Astronomy, Johns Hopkins University, Baltimore, MD 21218, USA}

    \author[0000-0002-6124-1196]{Yukei S. Murakami}
    \affiliation{Department of Physics and Astronomy, Johns Hopkins University, Baltimore, MD 21218, USA}
        
    \author[0000-0001-8596-4746]{Erik R.~Peterson}
    \affiliation{Department of Physics, Duke University, Durham, NC 27708, USA}

    \author[0000-0001-5201-8374]{Dillon Brout}
    \affiliation{Departments of Astronomy and Physics, Boston University, Boston, MA 02140, USA}

    \author[0000-0002-4934-5849]{Maria Acevedo}
    \affiliation{Department of Physics, Duke University, Durham, NC 27708, USA}

    \author[0000-0001-8596-4746]{Bastien Carreres}
    \affiliation{Department of Physics, Duke University, Durham, NC 27708, USA}

    \author[0000-0002-6230-0151]{David O.~Jones}
    \affiliation{Institute for Astronomy, University of Hawai`i, 640 N.~A'ohoku Pl., Hilo, HI 96720, USA}

   \author[0000-0002-1809-6325]{Khaled Said}
    \affiliation{School of Mathematics and Physics, University of Queensland, Brisbane, QLD 4072, Australia}
    \affiliation{OzGrav:~The ARC Centre of Excellence for Gravitational Wave Discovery, Hawthorn, VIC 3122, Australia}
 
    \author[0000-0002-1081-9410]{Cullan Howlett}
    \affiliation{School of Mathematics and Physics, University of Queensland, Brisbane, QLD 4072, Australia}
    \affiliation{OzGrav:~The ARC Centre of Excellence for Gravitational Wave Discovery, Hawthorn, VIC 3122, Australia}

    \author[0000-0002-5259-2314]{Gagandeep S.~Anand}
\affiliation{Space Telescope Science Institute, 3700 San Martin Drive, Baltimore, MD 21218, USA}

\begin{abstract}

The Dark Energy Spectroscopic Instrument (DESI) collaboration measured a tight relation between the Hubble constant ($H_0$) and the distance to the Coma cluster using the fundamental plane (FP) relation of the deepest, most homogeneous sample of early-type galaxies. To determine $H_0$, we measure the distance to Coma by several independent routes each with its own geometric reference.  We measure the most precise distance to Coma from 12 Type Ia Supernovae (SNe Ia) in the cluster with mean standardized brightness of $m_B^0=15.712\pm0.041$ mag. Calibrating the absolute magnitude of SNe~Ia with the {\it HST} distance ladder yields $D_{\textrm Coma}=98.5\pm2.2$ Mpc, consistent with its canonical value of 95--100 Mpc.   This distance results in $H_0=76.5 \pm 2.2$ km/s/Mpc from the DESI FP relation.  Inverting the DESI relation by calibrating it instead to the Planck+$\Lambda$CDM value of $H_0=67.4$ km/s/Mpc implies a much greater distance to Coma, $D_{\textrm Coma}=111.8\pm1.8$ Mpc, $4.6\sigma$ beyond a joint, direct measure. 
Independent of SNe~Ia, the {\it HST} Key Project FP relation as calibrated by Cepheids, Tip of the Red Giant Branch from {\it JWST}, or {\it HST} NIR surface brightness fluctuations all yield $D_{\textrm Coma}<$ 100 Mpc, in joint tension themselves with the Planck-calibrated route at $>3\sigma$.   From a broad array of distance estimates compiled back to 1990, it is hard to see how Coma could be located as far as the Planck+$\Lambda$CDM expectation of $>$110 Mpc. By extending the Hubble diagram to Coma, a well-studied location in our own backyard whose distance was in good accord well before the Hubble Tension, DESI  indicates a more pervasive conflict between our knowledge of local distances and cosmological expectations.  We expect future programs to refine the distance to Coma and nearer clusters to help illuminate this new, local window on the Hubble Tension.

\end{abstract}
\keywords{}

\section{Introduction}

The `Hubble Tension' refers to the discrepancy in the value of the Hubble constant, $H_0$, between multiple measures of local distance and redshift (clustering around $H_0\sim73$ km/s/Mpc) versus an inferred value based on measurements of the Cosmic Microwave Background and the standard model of cosmology (found to be around $H_0\sim67.5$ km/s/Mpc); see \cite{Verde23} for a review.  As there is not yet an accepted theory of new physics to explain this discrepancy, there has been a wide focus on new and improved ways to study this phenomenon.  

Recently, the Dark Energy Spectroscopic Instrument (DESI) collaboration (\citealp{Said24}, hereafter S24) 
measured a tight relation between the Hubble constant ($H_0$) and the distance to the Coma cluster using the fundamental plane (FP) relation of the deepest, most homogeneous sample of early-type galaxies.  The FP is a long-known relation for early-type galaxies between their velocity dispersion, surface brightness and apparent radius \cite{Djorgovski:1987,Dressler:1987} which adds a parameter and tightens the earlier Faber-Jackson relation between their velocity and luminosity \cite{Faber:1976}.
The DESI measurement consists of redshifts and uncalibrated FP distances to 4,191 early-type galaxies in the Hubble flow and 226 such FP distances in the Coma cluster.  DESI measures the Hubble flow at $0.023 < z < 0.1$ and Coma serves only as a reference location, rich in early type galaxies, where the uncalibrated FP distances may be calibrated from knowledge of the distance to Coma.  S24 found $H_0=76.05\pm0.35$ (statistical) $\pm0.49$ (systematic FP) $\pm4.86$ (FP calibration) km/s/Mpc, a result usefully described as, 

\begin{equation}
H_0=(76.05 \pm 1.3)*({99.1~\textrm{Mpc} \over D_{\textrm{Coma}}})~\textrm{km/s/Mpc}.
\label{eq:H0coma}
\end{equation}

\noindent The DESI FP-relation estimate of $H_0$ depends on knowledge of the distance to Coma which was obtained by S24 from a surface brightness fluctuation (SBF) measurement of one galaxy in Coma, $D_{\textrm{Coma}}=99.1\pm5.8$ Mpc (\citealp{Jensen21}; hereafter J21).  The uncertainty from the DESI measurement is modest at $\pm 1.3$ km/s/Mpc (dominated by the FP-relation measurement of 226 galaxies in Coma) so that the uncertainty in the DESI estimate of $H_0$ is dominated by knowledge of the distance to Coma, one of the richest nearby clusters in the local Universe ($z=0.023$; \citealp{DOnofrio97}).  Our paper attempts to improve on this uncertainty by measuring the distance to Coma with a new sample of a dozen Type Ia supernovae (SNe Ia) in the cluster and by leveraging other distance measurements from the {\it Hubble Space Telescope} ({\it HST}) and the {\it James Webb Space Telescope} ({\it JWST}) to improve the $H_0$ constraint.

Coma has a long history of distance measurements to the objects within it.  A historical compilation of distance measurements was presented in \cite{degrijs:2020}\footnote{Two mean distances are provided by \cite{degrijs:2020}  for Coma, 99.5 Mpc by absolute measures or 90.4 Mpc for measures relative to Virgo.  Importantly we note they cite only the dispersion of all measures rather than the error in their mean which would require an analysis of their correlated terms which the authors do not undertake.  The error in the mean is also needed for comparisons so we only cite their mean here.} and from the 1990's-2000's in \cite{Carter08} including the use of Tully-Fisher, SBF, $D_n-\sigma$, FP and Globular Clusters, resulting in an average of $D_{\textrm{Coma}}\sim$95 Mpc.  The {\it HST} Key Project calibrated the FP relation in nearby clusters (Virgo, Fornax and Leo I) and Coma which resulted in a measured distance of $86\pm8$ Mpc \citep{Kelson00,Freedman01,degrijs:2020}.
Coma is generally too far to reach directly with primary distance indicators (i.e., Cepheids, Tip of the Red Giant Branch (TRGB) stars, Miras, J-region Asymptotic Giant Branch (JAGB) stars, Blue Supergiants (BSG) stars), but is rich in early-type galaxies which are ideal targets for SBF or galaxy-based methods like the FP relation.

SNe Ia offer an especially capable tool for calibrating the distance to the Coma cluster.  
With canonical rates of 1 SN per galaxy per 100 years, we should expect on the order of $\sim$10s of SNe in Coma to have been discovered by various surveys over the last decade.   The most recent effort to collect SNe~Ia in Coma came from a 1990 study, `Distances of the Virgo and Coma Clusters of Galaxies through Novae and Supernovae' \citep{Capaccili90}, which gathered 5 from the 1960's and 70's going back to \cite{Zwicky:1961}, however the quality of this data by modern standards is quite poor.   A precise distance measurement requires  multiple SNe Ia whose light curves and spectra pass contemporary quality cuts from photometric systems that are well characterized with additional samples in the Hubble flow for empirical testing. Until the last decade, Coma was not continuously searched for transients, so many past SNe Ia would have been missed. The Pantheon$+$ compilation \citep{Scolnic22,Brout22} contains only 2 SNe Ia located in Coma. Recent surveys like the Asteroid Terrestrial-impact Last Alert System (ATLAS; \citealp{Tonry18, Smith20}) and the Zwicky Transient Facility (ZTF; \citealp{Bellm19}) cover large fractions of the northern sky. This includes the area of the Coma cluster at RA=13h and Dec=28$^{\circ}$, and queries of their databases indicate they have found $>$10 SNe Ia around Coma in the last few years.

In Section~\ref{sec:data}, we search for SNe located within the Coma cluster with publicly available catalogues and light curves.  We then fit the light curves in Section~\ref{sec:fit} to derive standardized brightnesses of the SNe. In Section~\ref{sec:distance}, we measure a precise distance to Coma based on the calibrations of the absolute luminosity of SNe~Ia. Combining with other methods independent of SNe~Ia, we also measure $H_0$ using the DESI measurements of S24. Our Discussion and Conclusions are given in Section~\ref{sec:discuss} and Section~\ref{sec:conclusion}.

\begin{figure*}
    \centering
    \includegraphics[width=2\columnwidth]{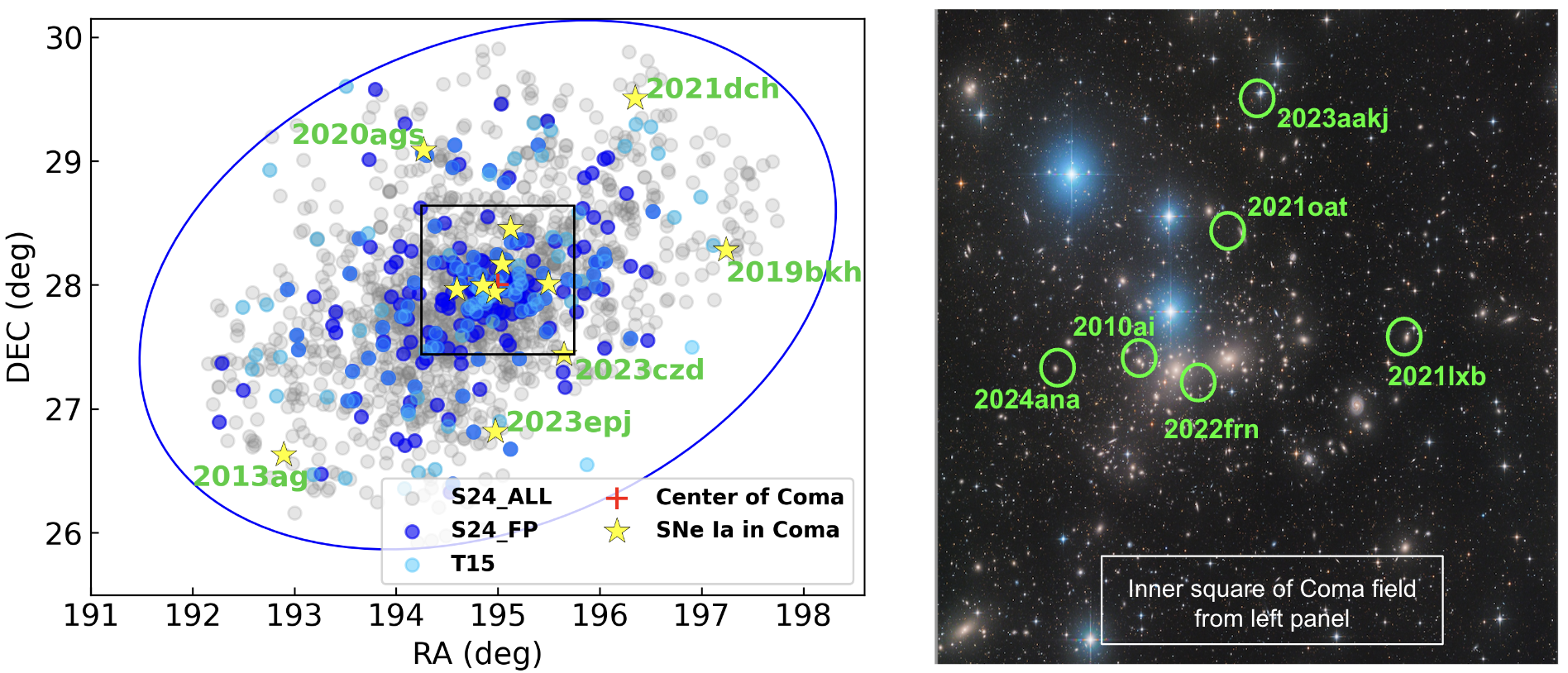}
    \caption{(Left) The locations of the SNe Ia identified to be in the Coma cluster (yellow star) and the galaxies identified to be in the Coma group as from the full S24 Coma group catalog (light gray), the S24 FP sample (dark blue) and the T15 Coma group catalog (light blue).  The center of the cluster is marked in red. The positions of the SNe are listed in Table~\ref{tab:tripp}. (Right) For the rectangular box on the left, a colored image of that sky area with the SNe within that location marked.}
    \label{fig:coma}
\end{figure*}

\renewcommand{\arraystretch}{1.0} 
\begin{deluxetable*}{lllllrrrrr}
\tablecaption{Properties of SNe Ia in Coma and their Light Curves \label{tab:tripp}}
\tablewidth{0pt}
\tablehead{
\colhead{SN} & $z_{hel}^{~~(\textrm a)}$ & P-SN$^{~~(\textrm b)}$ & P-Host$^{~~(\textrm b)}$ &  \colhead{$m_B$} & \colhead{$x_1$} & \colhead{$c$} & \colhead{Host Log-Mass} & \colhead{$\delta_B$} & \colhead{$m_B^0$} 
}
\startdata
\textbf{~~~ATLAS SNe} & ~ & ~ & ~ & ~ & ~ & ~ \\
2019bkh & 0.0195 & \makecell{197.2391 \\ 28.2812} & \makecell{197.2381 \\ 28.2804} & $15.96 \pm 0.03$ &  $0.76 \pm 0.07$ &  $0.09 \pm 0.02$ &  $9.40$ &  $0.04$ & $15.76 \pm 0.17$ \\
\hline
2020ags & 0.0206 & \makecell{194.2706 \\ 29.0887} & $~~~---~~~$$^{(c)}$ & $15.46 \pm 0.02$ &  $0.09 \pm 0.07$ & $-0.06 \pm 0.01$ &  $7.80$ &  $0.05$ & $15.60 \pm 0.12$ \\
\hline
2021dch & 0.0202 & \makecell{196.34900\\ 29.5096} & \makecell{196.3484 \\ 29.5104} & $15.99 \pm 0.03$ & $-1.42 \pm 0.06$ & $-0.08 \pm 0.03$ &  $9.55$ &  $0.08$ & $15.94 \pm 0.13$ \\
\hline
2021lxb$^{~~(\textrm d)}$ & 0.0259 & \makecell{195.4944 \\ 28.0086} & \makecell{195.4900 \\ 28.0058} & $16.75 \pm 0.05$ & $-2.13 \pm 0.08$ &  $0.22 \pm 0.03$ & $11.07$ & $-0.12$ & $15.86 \pm 0.28$ \\
\hline
2022frn & 0.0230 & \makecell{194.9657 \\ 27.9435} & $~~~---~~~$$^{(c)}$ & $15.93 \pm 0.02$ & $-1.28 \pm 0.07$ & $-0.02 \pm 0.02$ &  $7.51$ &  $0.04$ & $15.76 \pm 0.13$ \\
\hline
2023aakj & 0.0241 & \makecell{195.1214 \\ 28.4555} & \makecell{195.1223 \\ 28.4557} & $15.66 \pm 0.07$ & $-1.52 \pm 0.06$ & $-0.21 \pm 0.04$ &  $9.13$ &  $0.22$ & $15.86 \pm 0.15$ \\
\hline
2023czd & 0.0180 & \makecell{195.6467 \\ 27.4393} & \makecell{195.6465 \\ 27.4394} & $15.31 \pm 0.03$ &  $0.36 \pm 0.06$ & $-0.08 \pm 0.02$ &  $9.50$ &  $0.07$ & $15.54 \pm 0.11$ \\
\hline
2023epj & 0.0267 & \makecell{194.9738 \\ 26.8194} & \makecell{194.9764 \\ 26.8200} & $15.36 \pm 0.01$ &  $0.33 \pm 0.06$ & $-0.12 \pm 0.01$ &  $9.62$ &  $0.12$ & $15.67 \pm 0.12$ \\
\hline
2024ana & 0.0201 & \makecell{194.5941 \\ 27.9662} & \makecell{194.5908 \\ 27.9678} & $16.31 \pm 0.03$ & $-2.44 \pm 0.08$ &  $0.07 \pm 0.02$ &  $9.97$ & $-0.00$ & $15.75 \pm 0.17$ \\
\hline
\textbf{~~~YSE SNe} & ~ & ~ & ~ & ~ & ~ & ~ \\
2021lxb$^{~~(\textrm d)}$ & 0.0259 & \makecell{195.4944 \\ 28.0086} & \makecell{195.4900 \\ 28.0058} & $16.58 \pm 0.04$ & $-2.29 \pm 0.04$ &  $0.14 \pm 0.03$ & $11.07$ & $-0.05$ & $15.84 \pm 0.22$ \\
\hline
2021oat & 0.0225 & \makecell{195.0344 \\ 28.1703} & \makecell{195.0378 \\ 28.1704} & $15.22 \pm 0.04$ & $-0.04 \pm 0.07$ & $-0.18 \pm 0.03$ & $10.19$ &  $0.14$ & $15.65 \pm 0.14$ \\
\hline
\textbf{~~~Pantheon$+$ SNe} & ~ & ~ & ~ & ~ & ~ & ~ \\
2010ai & 0.0183 & \makecell{194.8501 \\ 27.9964} & \makecell{194.8544 \\ 27.9967} & $15.68 \pm 0.04$ & $-1.65 \pm 0.09$ & $-0.12 \pm 0.03$ & $9.11$ & $0.08$ & $15.73 \pm 0.10$ \\
\hline
2013ag & 0.0213 & \makecell{192.8959 \\ 26.6293} & \makecell{192.8987 \\ 26.6295} & $15.80 \pm 0.04$ & $-1.11 \pm 0.75$ &  $0.02 \pm 0.06$ & $8.79$ & $0.01$ & $15.57 \pm 0.21$ \\
\hline
\hline
\enddata
\tablecomments{(a) The heliocentric redshifts are all from the host galaxy spectra ($\sigma_z=0.0001$) except in the cases of 2020ags and 2022frn, which do not have obvious hosts. The redshift for these SNe Ia comes from the SN spectra themselves and $\sigma_z=0.005$. (b) Positions of the SN and host galaxy, respectively. (c) No host associated within 15''. (d) 2021lxb is the only SN with light curves from two samples.}
\end{deluxetable*}

\section{Data}\label{sec:data}

%I dont want no scrubs, a scrub is a guy who cant get no love from me

Given our goal of measuring a reliable distance to Coma, we collect SNe Ia whose membership in Coma can be established with high confidence by their 3-dimensional coordinates and with data suitable for high quality distance measurements.  In order to ascertain whether SNe Ia may be located in the Coma cluster, we first determine the size and location of the cluster from a sample of virially-associated galaxies. Following \citet{Peterson22}, we query the `group' catalog that defines groups of galaxies in the nearby universe ($z<0.04$) from \citet{Tully15} (hereafter T15), as well as the new full Coma group catalog from S24 and the subset of these galaxies used in S24 to determine FP measurements.  We show the galaxy members according to these three subsamples in Fig.~\ref{fig:coma} (left).  The catalogs share a common boundary and a common redshift range of $0.015<z<0.032$, both with a median heliocentric redshift of $\sim0.0232$ and standard deviation of $0.003$. The center (RA, DEC) for both catalogs is ($\sim195^{\circ}$, $\sim28^{\circ}$).

From T15, the Coma group contains 148 members and from the full DESI Coma group, the group contains 1696 members.  T15 is limited to brighter, larger members, but they are sufficient as a crosscheck for defining the inner bound region of Coma with high confidence. The incompleteness from T15, as discussed in \cite{Peterson24b}, is due to the magnitude-limited galaxy surveys used to make up the group catalogs.  These catalogs provide the redshift range of the group as listed above, as well as the spatial contours in RA and DEC, which are presented in Fig.~\ref{fig:coma}.  We find an ellipse with center ($195^{\circ}$, $28^{\circ}$) with semi-major RA axis of 3.5 degrees, semi-minor DEC axis of 2 degrees, and angle of $15^{\circ}$ encloses all of the galaxy members of these catalogs. The right panel of Fig.~\ref{fig:coma} highlights the region with the highest concentration of galaxies within the cluster.\footnote{\url{https://www.astrobin.com/full/fh0dto/C/}.}

Given the constraints in redshift and location from the group catalog, we query the Transient Name Server (TNS\footnote{\url{https://www.wis-tns.org/}.}) and SIMBAD \citep{simbad} for spectroscopically-confirmed SNe Ia.  We query for SNe post-1990 from multiple servers, though note that TNS appears to be incomplete for years before 2015, and SIMBAD is incomplete for years after 2020. We find 32 spectroscopically-confirmed SNe Ia within the ellipse as shown in Fig.~\ref{fig:coma} and report their names, positions, host-galaxy positions and heliocentric redshifts in the Appendix.  The majority of SNe discovered after 2018 were found/measured by ATLAS and ZTF, as well as a smaller amount by the Young Supernova Experiment (YSE, \citealp{Jones21}; public data release \citep{Aleo23}).  We retain all spectroscopically-normal SNe~Ia with high quality light curves following well-established quality criteria \citep{Scolnic15}, see Appendix for further detail. All SNe included in our main analysis are associated to a host galaxy in the S24 full group catalog with the exception of 2020ags and 2022frn, which appear to be `hostless'.  Still, the redshifts of these two SNe are within the nominal redshift range for Coma, and within the ellipse outlined above.  

\begin{figure}
    \centering
    \includegraphics[width=1.0\columnwidth]{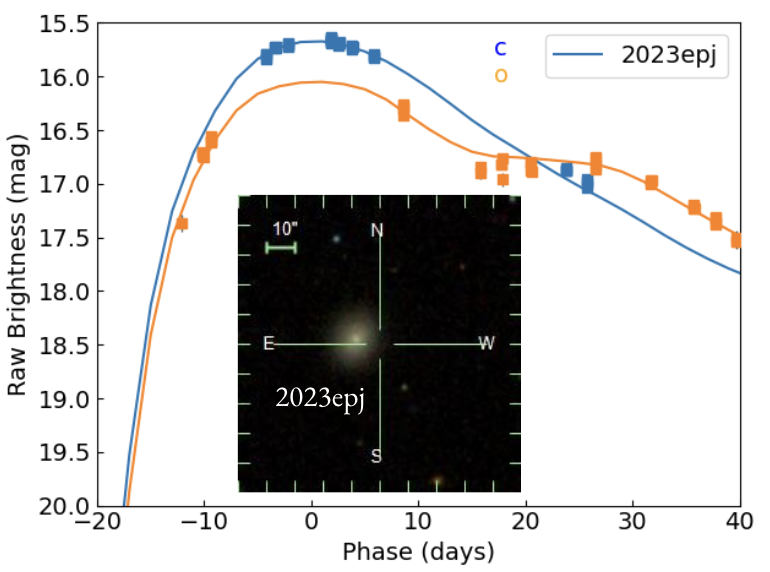}
    \caption{ Raw light curve data in the `c' and `o' bands from the ATLAS survey of one representative SN: 2023epj.  The curves are the SALT2 model fit. The inset shows the position of the SN relative to its host galaxy \citep{sdss09}  This figure for each SN Ia, along with the raw data, is online at \url{https://github.com/dscolnic/Coma}.
    }
    \label{fig:lc}
\end{figure}

We provide an example light curve from the ATLAS database for a SN Ia in Coma in Fig.~\ref{fig:lc}; we also show the stamp of the host galaxy of the SN.  We share all the light-curve data and figures like Fig.~\ref{fig:coma} for each SN on \url{https://github.com/dscolnic/Coma}.
We find two SNe already in the Pantheon+ compilation that are located in the cluster: 2010ai and 2013ag. Both ATLAS and YSE photometry are photometrically calibrated to the Pan-STARRS \citep{Chambers16} AB magnitude system.  YSE photometry was measured with the Pan-STARRS telescope \citep{Chambers16}, and thus previous system characterization (e.g.,~\citealp{Foley18,Scolnic22}) can be used.  Since the ATLAS and YSE photometric systems are relatively new and less studied, we undertake an additional procedure, remeasuring the Hubble flow with each, to check for survey offsets in the following section.

\section{Measuring a Mean Standardized Brightness for SNe Ia in the Coma Cluster}\label{sec:fit}

\subsection{Light-curve Fitting and Standardization}

We use the SALT2 light-curve model (introduced in \citealp{Guy10}; modified in \citealp{Brout22}) to be consistent with past $H_0$ measurements as described in the Pantheon+SH0ES analysis \citep{Brout22}.  We apply light-curve quality requirements as described in \cite{Scolnic22}, which can be summarized as a SALT2 color $c$ within ($-0.3,0.3$); a SALT2 stretch $x_1$ within ($-3,3$); measurements in two filters with a signal-to-noise ratio $>5$; at least one measurement before peak and spectroscopically classified as a normal SN Ia (following Pantheon+, including SN Ia-91T, excluding SN Ia-91bg).  Of the 32 SNe from the literature search, there are 12 SNe with light curves that  are publicly available and pass the SALT2 fit and the quality checks.  One, 2021lxb, was observed by both ATLAS and YSE.  The most common reason for failure was due to inadequate coverage from multiple bands.  Explanation of why specific SN~Ia did not pass cuts is included in the Appendix.   

We show the positions of all the SNe Ia that pass cuts in Fig.~\ref{fig:coma}. We present the SALT2 fit parameters ($m_B, x_1, c$) in Table~\ref{tab:tripp}. In Pantheon+, the standardized brightness $m_B^0$ was calculated by, 
\begin{equation}
m_B^0 = m_B+0.15x_1-3.1c-\delta_B, 
\label{eq:Tripp}
\end{equation}
where the multiplicative coefficients for stretch and color were determined in Pantheon+ to minimize the Hubble residual dispersion.

The term, $\delta_B$,  known as a `bias-correction,' is a standard measure of the difference between the the population mean and the statistical selection of a SN Ia sample as predicted from simulations of discovery and follow-up \citep{KesslerScolnic17}.  More recently this includes a function which accounts for (i.e., simulates) the empirical correlation between standardized SN Ia brightness and host-galaxy mass \citep{BroutScolnic21}.   Hosts masses are given in Table~\ref{tab:tripp} and are derived for this sample following the approximation determined in \cite{Taylor11} and photometry from \cite{Chambers16}.   To measure bias corrections following \cite{KesslerScolnic17}, we simulate the ATLAS and YSE samples with the SNANA package \citep{Kessler09}.  We generate properties of the survey directly from the light-curve data of the respective samples and determine a selection function so that the redshift distribution between the simulation and data match.  We follow the simulation methodology in \cite{Brout22} and use the intrinsic scatter model developed in \cite{Popovic21} that describes observed color as a mix of intrinsic color, dust reddening, and noise. The uncertainty on the magnitudes is determined following \cite{KesslerScolnic17} and is dependent on the fitted SALT2 $c$ and $x_1$ of each SN.   The mean of the bias corrections is larger than average samples, $0.048\pm0.021$ mag, due to 6 of the 13 light curves having $c\leq - 0.08$. These less-dusty, blue SNe are more common in ellipticals common to clusters \citep{Chen22} and standardization is biased for the bluest SNe due to the use of a single color-luminosity standardization coefficient (i.e. 3.1 in Eq.~\ref{eq:Tripp}) as discussed in \cite{Scolnic16}. If we split the sample in half, sorted by the bias correction size, the offset in $m_B^0$ between the two subsamples is $0.0\pm0.08$ mag.

 When we compare 38 SNe common to both the full samples of YSE and ATLAS SNe (i.e., not limited to Coma), we find a difference in SN color of 0.04 mag.  This is not surprising as the ATLAS `co' bands are novel and different than the common $griz$ (YSE) or $BVRI$ filters used and well-studied for most SN Ia samples.  We attribute this difference to a modest mis-estimate of the preliminary calibration of the ATLAS `c' band, a finding further indicated by comparing the ATLAS color distributions between simulations and data, and add the 0.04 mag to the ATLAS `c'-band photometry.

\begin{figure*}[!hbt]
    \centering
    \includegraphics[width=2\columnwidth]{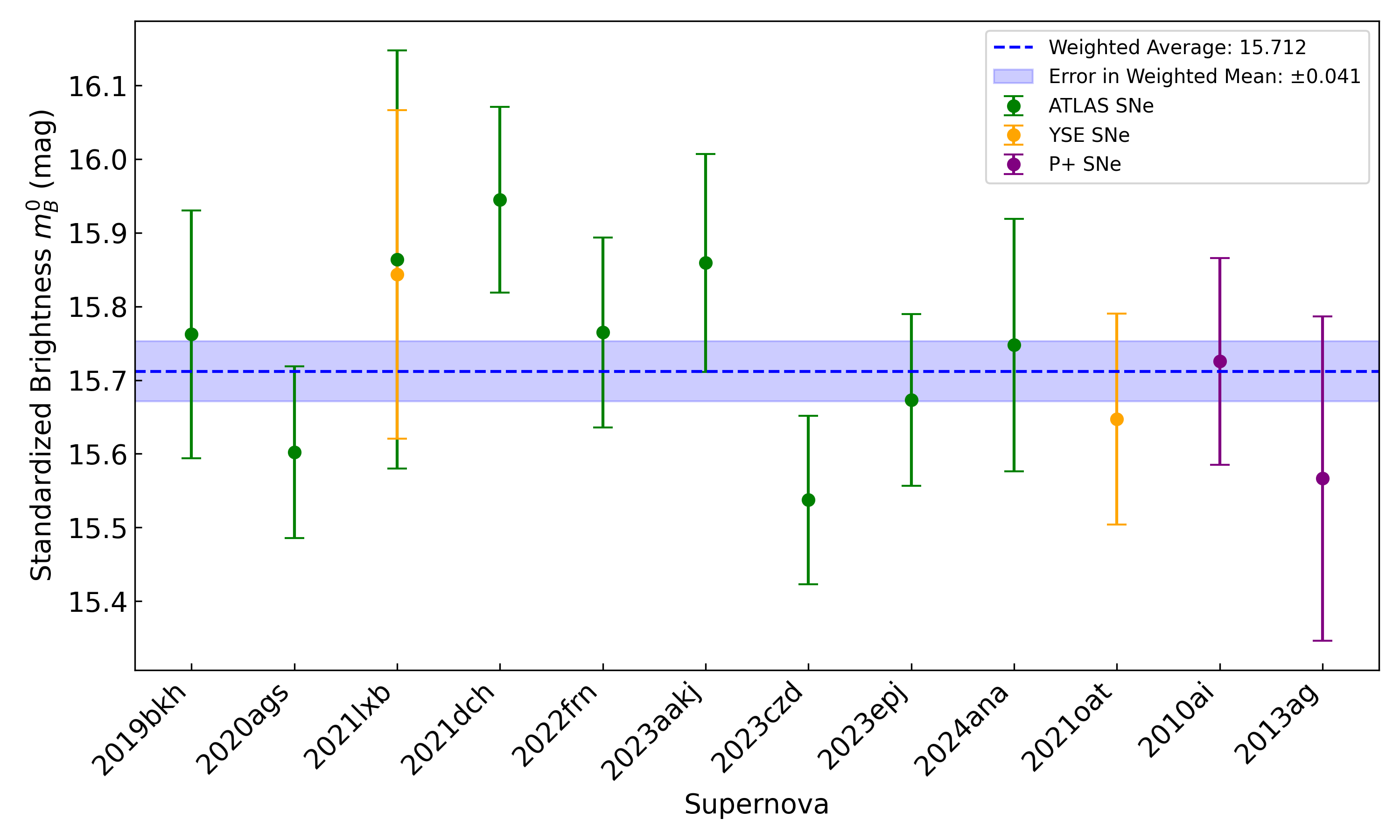}
    \caption{Standardized magnitudes for SNe Ia in the Coma cluster whose light curves pass quality cuts.  The average line is shown in dashed-blue.  The colors of the SNe are tied to the survey sample.
}
    \label{fig:whisker}
\end{figure*}

We present all the standardized brightness values of the SNe in Coma in Table~\ref{tab:tripp} and Fig.~\ref{fig:whisker}. We find very good agreement for the standardized brightnesses measured for the SNe in and around the cluster.  The range of magnitudes for the 12 SNe is 15.54 mag to 15.94 mag. The SN, 2021lxb, which is common to both YSE and ATLAS samples, has a difference in standardized brightness between the two of 0.02 mag; we therefore use the light curve from YSE as the standardized brightness has smaller uncertainty.  Weighting for the individual uncertainties, we measure an average standardized brightness derived from Eq.~\ref{eq:Tripp} for 12 SNe Ia in Coma to be: 

\begin{equation}
\label{eq:result}
\overline{m_B^0}=15.712\pm0.041 \textrm{ mag}.
\end{equation}

\subsection{Checking Accuracy of Measurements}

First, we check that the calibration of the YSE and ATLAS samples are consistent with samples used in Pantheon+ (so that we can tie to the absolute calibration determined there).  We follow the methodology in \cite{Brownsberger} to check whether mean standardized SNe~Ia magnitudes in different samples produce the same magnitude at a given redshift.  This is accomplished by comparing the calculated Hubble diagram intercept ($a_B$) for ATLAS and YSE in bins to that found for the full sample used in \cite{Riess22} of $a_B=0.714\pm0.002$.  The $a_B$ calculation is:
\begin{equation}
\begin{split}
 a_B = \log \left( cz \left[ 1 + \frac{1}{2}(1 - q_0)z 
 - \frac{1}{6}(1 - q_0 - 3q_0^2 + j_0)z^2 \right] \right) \\
 + O(z^3) - 0.2 {m_B^0}.
\end{split}
\label{eq:ax}
\end{equation}

We note that $a_B$ need not be used to solve for survey offsets, but is an effective summarization of the mean magnitude of a sample. We find very good agreement (at $a_B$ level of 0.002) inferred for YSE and ATLAS compared to that from Pantheon+, implying remaining calibration differences on the $\sim0.01$ mag level, which is negligible compared to the statistical uncertainty when added in quadrature.

The total $\chi^2$ measured for  Fig.~\ref{fig:whisker} is $\chi^2=9.1$ for 12 individual SNe with a sample dispersion of 0.12 mag, similar to that found for SNe~Ia samples in the Hubble flow (e.g., Pantheon$+$ dispersion is 0.13 mag).  Monte Carlo simulations of the same number of SNe show a $\chi^2$ this low or lower should happen $\sim30\%$ of the time for 12 SNe. 

As discussed in the Appendix, two SNe, 1994S and 2020jhf, are both within the angular range but right at the low-redshift limit and are not included as they may be in front of the cluster. These two SNe are $1.0$ and $0.5$ mag brighter respectively than the mean.  Inclusion of these two SNe would reduce the Coma distance by $\sim$ 0.10 mag (and as discussed in the next section, increase $H_0$). Another supernova excluded from the sample, 2015M, is classified \citep{Hicken07} as super-Chandrasekhar \citep{Howell} and appears to be a $3\sigma$ outlier compared to the other brightnesses in Table~\ref{tab:tripp}.
There are two SNe --- 2019be and 2006bz, that would produce fainter mean values if included, but are cut by the $x_1$ requirement and are also classified as 1991bg-like \citep{1991bg}, a peculiar SNe~Ia subtype routinely excluded in cosmological samples \citep{Scolnic22}.  Because there is greater volume beyond Coma than in front of it, statistical inclusion of non-members would tend to over-estimate the standardized brightness.  Future, larger samples of SNe in Coma may benefit from a simultaneous analysis of membership and standardized brightness.  For this sample we note both the conservative membership and our uncertainty in the mean (in the sense that we used the larger, expected  errors rather than renormalizing them by the lower dispersion of the sample).

\begin{figure*}[!hbt]
    \centering
    \includegraphics[width=2.1\columnwidth]{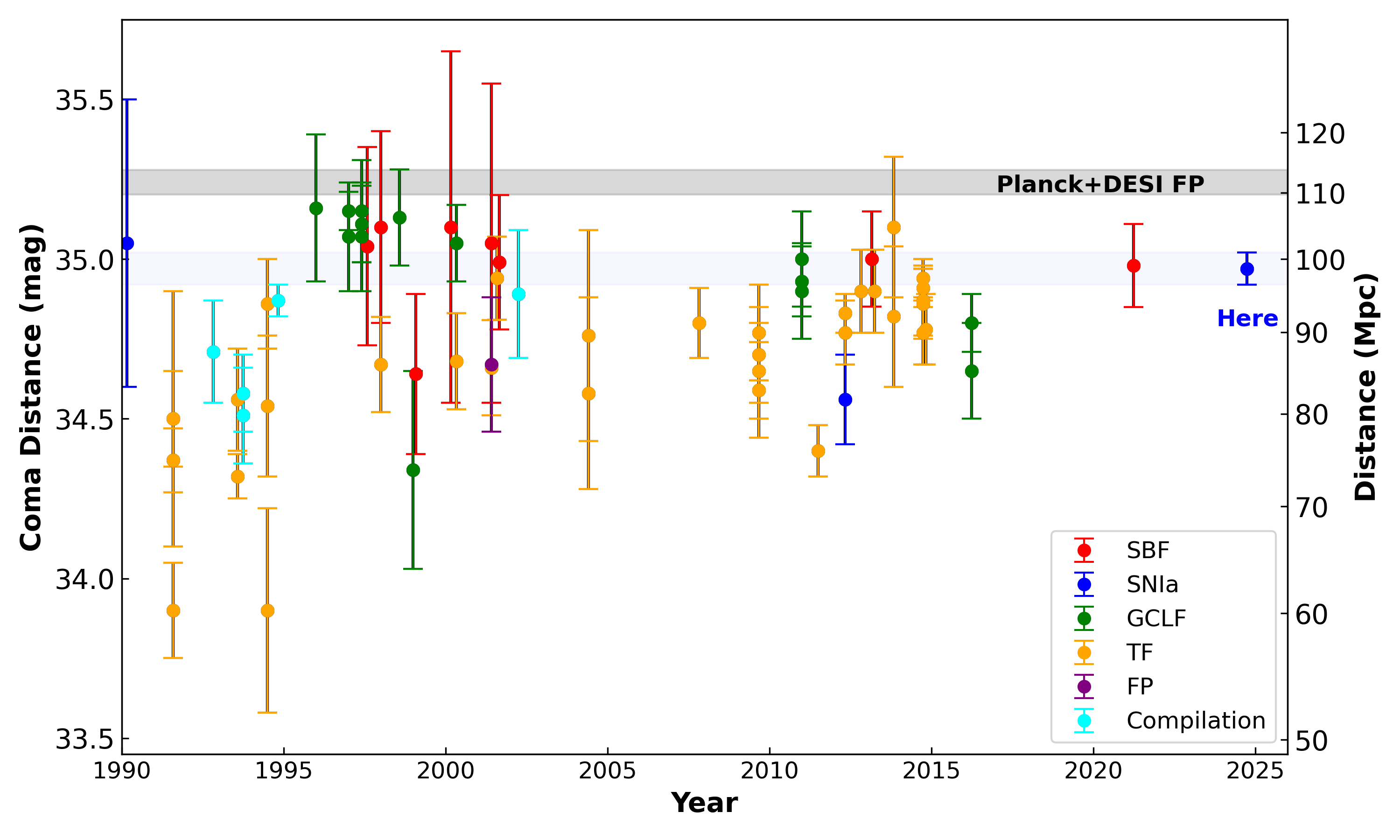}
    \caption{Historical (1990 onward) distance modulus measurements of the Coma cluster (as reviewed in \citealp{degrijs:2020}).  Only distance measurements that do not depend on redshift and $H_0$ are included.
}
    \label{fig:history}
\end{figure*}

We check to see if there is any relation between any of the properties of the SN Ia and the standardized brightness.  We calculate trends between $m_B^0$ and the following properties: redshift, separation between SN and host positions, separation from center of cluster, host mass, and the $x_1$ and $c$ fit parameters.  We do not find even mild evidence for a non-zero slope; the lowest p-value, expressing the chance that a recovered slope is consistent with zero, is only as low as 23\%.  We note that our sample does have a high relative fraction of fast-declining SNe (low $x_1$ values); this is likely due to the fact that more SNe are found in ellipticals in a cluster, and those SNe are more often fast-declining \citep{Sullivan10, Chen22}. We verify that fast-declining SNe Ia are standardized without significant bias by checking those in the Pantheon+ sample and find that the mean Hubble residuals for $x_1<-1.0$ is $+0.01\pm0.02$ mag.

% Add this line to increase the row spacing
\renewcommand{\arraystretch}{1.5}  % Adjust the value (e.g., 1.5) to increase/decrease spacing

As a crosscheck of the mean standardized brightness found here, we examine the Pantheon+ Hubble diagram to measure the mean standardized brightness for SNe at the redshift of the Coma cluster. We measure the Cosmic Microwave Background (CMB)-frame redshift of the Coma cluster from the group catalog as shown in Fig.~\ref{fig:coma} and find the mean to be 0.02422, and the median is $0.02445\pm 0.00024$. If we query the Pantheon+ sample for SNe with CMB redshifts within 0.005 of $0.02422$, we 
 find a mean magnitude of $m_{B,corr}=15.71\pm0.022$, in $<1\sigma$ agreement with that directly found for SNe in the Coma cluster.  This indicates that the mean redshift of the Coma cluster is accurately measured using the group-average and that calibration / bias-correction differences between the new samples and the Pantheon+ sample are likely limited to the $\sim0.03$ mag level.  We thus consider our mean result in Eq.~\ref{eq:result} as robust and representative of the mean standardized brightness of a SN Ia in the Coma cluster.

\begin{deluxetable*}{llll}
\tablecaption{\label{tab:Distances}The Distance to Coma Determined by 
Various Methods}
\tablehead{
\colhead{Technique}&\colhead{Distance (Mpc)}&\colhead{Distance Modulus (mag)}&\colhead{Reference}}
\startdata
\hline
{Literature Compilation \citep{Carter08}} \\
I-band Tully-Fisher&86.3$\pm$6&34.68$\pm$0.15&\cite{Tully00}\\
K$^{\prime}$-band SBF&85$\pm$10&34.64$\pm$0.27&\cite{Jensen99}\\
I-band SBF&102$\pm$14&35.04$\pm$0.32&\cite{Thomsen97}\\
$D_n-\sigma$&96$\pm$6&34.90$\pm$0.14&\cite{Gregg97}\\
FP &108$\pm$12&35.16$\pm$0.25&\cite{Hjorth97}\\
Globular Cluster Luminosity Function (GCLF) &102$\pm$6&35.05$\pm$0.12&\cite{Kave00}\\
\hline
{Literature Mean} & 95$\pm$3.1 & {34.89$\pm$0.06} &  \\
{{\it HST} KP FP} $^{~\textrm{(a)}}$ & 85$ \pm 8$ & 34.67$\pm 0.21$ & \cite{Kelson00,Freedman01} \\
\hline
{\bf JWST TRGB+FP} & 90$\pm$9 & 34.74$\pm$0.18 & \cite{Anand24a} \\
{\bf {\it HST}~NIR SBF} & 99.1$\pm5.8$ & 34.98$\pm0.13$ & J21 \\
{\bf SH0ES Cepheids+SNe~Ia} & 98.5$\pm 2.2$ & 34.97$\pm0.05$ & Here \\
\hline
{\bf Combination of Independent} & {\bf 98.0$\pm {\bf 2.0}$} & {\bf 34.965}$\pm{\bf 0.045}$ & Here \\
\hline
\enddata 
\tablecomments{(a) See \cite{degrijs:2020} which updates the {\it HST}~KP Cepheid calibration of LMC $18.50 \pm 0.13$ with the DEB result of 18.477 $\pm 0.026$ mag from \cite{Pietr19}.} 
\label{tab:lit}
\end{deluxetable*}

\section{The Distance to Coma}\label{sec:distance}

\subsection{Converting SN Ia Brightness into Distance}

The luminosities of standardized SNe~Ia, $M_B$, have been calibrated by other standard candles (e.g., Cepheids or TRGB, themselves calibrated geometrically) so that we can measure the distance (modulus) to Coma,
\begin{equation}
\label{eq:distance}
\mu = m_B^0 - M_B .
\end{equation} The most precise measurement of $M_B$ for SNe~Ia (and with Pantheon$+$ standardization) comes from the calibration of 42 SNe~Ia with measurements of {\it HST} Cepheids and 4 geometric anchors (\citealp{Riess22}, hereafter R22).  As shown in \cite{Riess24} and with measurements from \cite{Freedman24} for the largest JWST samples, {\it HST} Cepheids yield consistent distances (to within $\sim$ 1$\sigma$) with 8 other methods or telescope samples; {\it JWST} Cepheids, TRGB, and JAGB by two groups, plus {\it HST} TRGB and Miras.   We take the baseline value from R22 of $M=-19.253\pm 0.027$ mag, then following Eq.~\ref{eq:distance}, we measure a distance modulus to Coma of $34.97 \pm 0.05$ mag. This can be directly converted into a distance to Coma following,
\begin{equation}
\textrm{Distance (Mpc)} = 10^{0.2\times (\mu-25)}
\end{equation}
resulting in $D_{\textrm{Coma}}=98.5 \pm 2.2$ Mpc. We note that this is significantly closer, by 4.2$\sigma$, than the expected result of $\mu=35.24 \pm 0.039$ or $D_{\textrm{Coma}}=111.8\pm$1.8 Mpc if the DESI Hubble relation is instead calibrated with $H_0=67.4 \pm 0.5$ km/s/Mpc from Planck+$\Lambda$CDM (see Table 3 for relevant uncertainty terms).
This distance results in $H_0=76.5 \pm 2.2$ km/s/Mpc from the DESI FP relation. 
The difference between this and the result from the HST distance ladder, $H_0=73.0 \pm 1.0$ km/s/Mpc, removing error terms in common, is $3.5 \pm 2.3$ km/s/Mpc, 1.5$\sigma$.

\begin{figure*}[!hbt]
    \centering  \includegraphics[width=2.1\columnwidth]{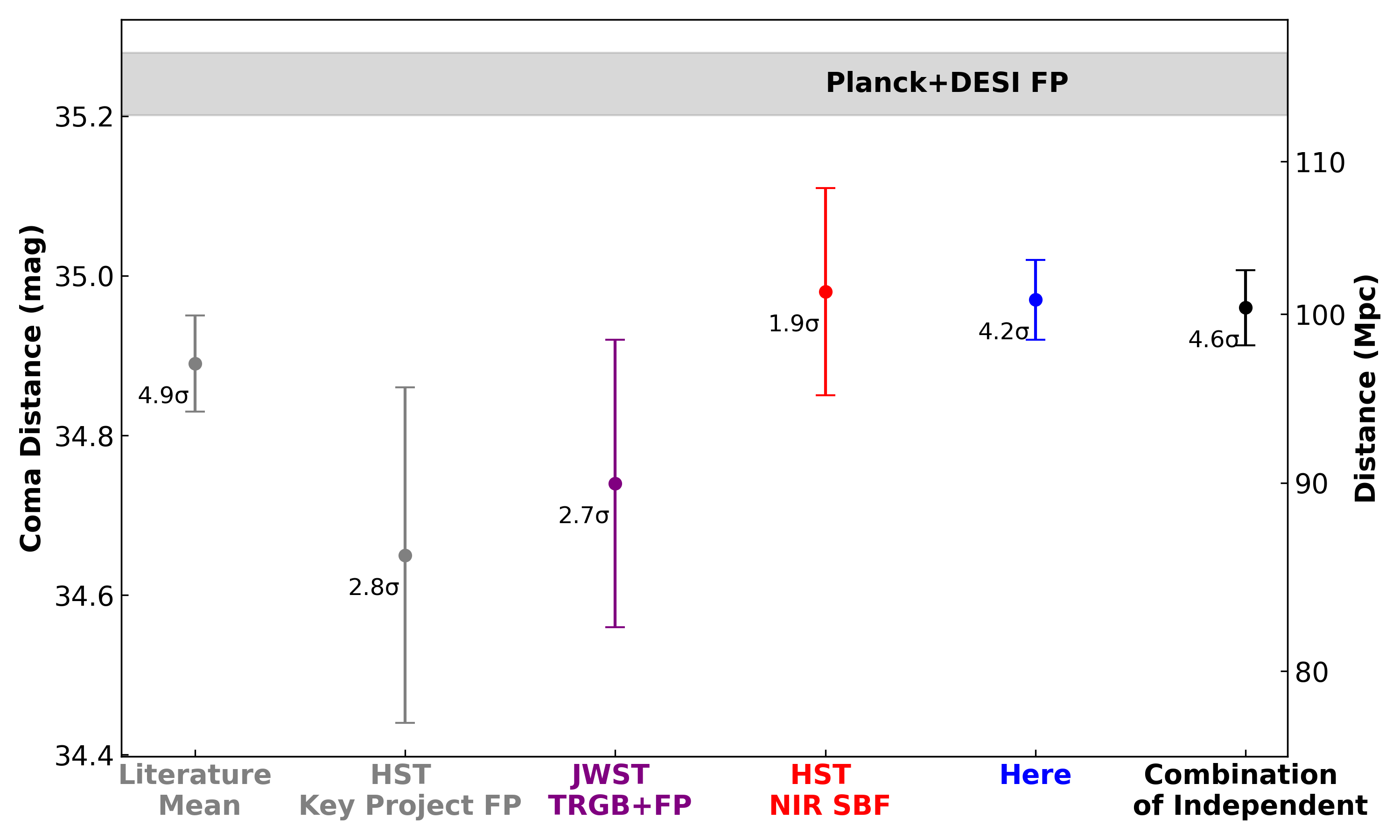}
    \caption{The distance modulus measurements to the Coma cluster as summarized in Table~\ref{tab:lit}. The `Combination of Independent' refers to combining JWST TRGB$+$FP, HST NIR SBF and the Cepheid$+$SN Ia measurement from this paper, all independent techniques (three colored points). 
}
    \label{fig:lit}
\end{figure*}

\subsection{Non-SN Distance to Coma}

There have been numerous studies of the distance to Coma.  An extensive review of more than 60 distance measurements to Coma going back more than 5 decades is presented in \cite{degrijs:2020}, see Fig.~\ref{fig:history}.   We exclude those based on the use of redshift and assumption for a value of the Hubble constant to avoid circularity.  These distance estimates demonstrate a tightening of the range in the last twenty years. We narrow the focus to a smaller compilation of more recent literature by \cite{Carter08} (their table 1) which gives distances from 6 studies: I-band Tully-Fisher \citep{Tully00}, K-band SBF \citep{Jensen99}, I-band SBF \citep{Thomsen97}, $D_n - \sigma$ \citep{Gregg97}, FP \citep{Hjorth97} and Globular Cluster Luminosity Function (GCLF, \citealp{Kave00}).  These have a mean of $D_{\textrm{Coma}}=95.1 \pm 3.1$ Mpc and are consistent as a set with a $\chi^2=5.9$ for the 6 values, indicating reasonable consensus.  
We do not attempt to determine the correlated terms in the \cite{Carter08} sample but rather defer that step to a combination of more recent measures in the next section.

The {\it HST} Key Project (KP) undertook a study of the FP for early-type galaxies in a range of nearby galaxy clusters and in Coma \citep{Kelson00}.  In principle this is a logical extension of the DESI FP measure of the Hubble flow calibrated to Coma, but to even nearer galaxy clusters via the same tool. By measuring the FP relations in 3 local clusters, Virgo, Fornax, and Leo I, with 26 galaxies, \cite{Kelson00} established a zeropoint for the FP, calibrated by the {\it HST} KP Cepheid data set and 81 galaxies in Coma from \cite{Jorgensen95b,Jorgensen95a}.   The random and systematic uncertainty in the FP is given by \cite{Kelson00} as 0.02 dex (5\% in distance) and 0.03 dex (7\% in distance) respectively.  The updated Cepheid zeropoint error including the LMC Detached Eclipsing Binary (DEB) distance is 4\% \citep{Pietr19}.  The result is an updated {\it HST} KP FP Coma distance estimate of $\mu=34.67 \pm 0.21$ mag or 85 $\pm$ 8 Mpc, including systematic uncertainties as summarized by \cite{degrijs:2020}.

The TRGB-SBF Project has recently measured with JWST the distance to the same 3 local clusters used by the {\it HST} KP to measure the Coma distance.  Using the values given in \cite{Anand24a,Anand24b} to recalibrate the {\it HST} KP FP relation results in $90 \pm 9$ Mpc.
Lastly we include in Table~\ref{tab:Distances} the {\it HST} near-infrared (NIR) SBF estimate of the Coma distance from J21 used by S24 (DESI), $99.1 \pm 5.8$ Mpc, based on a single galaxy.

\subsection{Combining Independent Techniques }

We can increase the precision of the measurement from an average of {\it uncorrelated} measurements from the most recent distances measurements to Coma.  For each measurement, we describe the calibration ladder to reach Coma. The J21 distance to Coma in Table~\ref{tab:Distances} is based on SBF which is calibrated by Cepheids measured by the {\it HST} KP, and these Cepheids are calibrated by the DEB distance to the LMC \citep{Pietr19}.  The JWST TRGB measurements from \cite{Anand24a,Anand24b} may be used to recalibrate the {\it HST} KP FP from the same 3 local clusters, Virgo, Fornax and Leo I and then follow the same FP link to Coma.
The JWST TRGB itself is calibrated by NGC 4258 \citep{Reid19}.  Lastly, for the combination analyzed in this paper, we may adopt the calibration of SNe~Ia by SH0ES Cepheids, which  are calibrated only by Gaia EDR3 Milky Way parallaxes (R22; fit 11 in their table 5).  

The distance constraint from combining these three independent measurements is $D_{\textrm{Coma}}=98.0 \pm 2.0$ Mpc or $\mu_0=34.957 \pm 0.045$ mag. 
 Using Eq.~\ref{eq:H0coma} converts this into $H_0=76.9 \pm 2.0$ km/s/Mpc.  We show these independent approaches as well as the combination in Fig.~\ref{fig:lit}.
 As also shown in Fig.~\ref{fig:lit}, this distance found from combining these local measurements is $\sim 4.6\sigma$ ($0.28 \pm 0.06$ mag) from the  the distance to Coma of $D=111.8\pm1.8$ Mpc 
if one combines the DESI FP Hubble relation with $H_0$ from Planck+$\Lambda$CDM of $67.4$ km/s/Mpc.

\begin{deluxetable*}{lll}
\tablecaption{\label{tab:errs } Coma and $H_0$ Uncertainties in a DESI-based Distance Ladder}
\tablehead{
\colhead{Source}&\colhead{Value (mag)}&\colhead{Reference}}
\startdata
SNe~Ia $M_B$ & 0.027 & \cite{Riess22} \\
Coma SNe~Ia $m_B$ & 0.041 & Here \\
\hline
Coma Distance subtotal & 0.049 & \\
\hline
FP in Coma (N=226) $^{(\textrm a)}$ & 0.033 &  \cite{Said24} \\
FP in Hubble Flow (N=4191)+DESI sys. & 0.017 & \cite{Said24} \\
\hline
\textbf{Local subtotal} ($H_0$) & 0.063 & \\
\hline
\hline
Planck+$\Lambda$CDM & 0.013 & \cite{Planck18}\\
\hline
\textbf{Comparison Total} & 0.064 $^{(\textrm b)}$  &  \\
\hline
\enddata 
\tablecomments{(a) Each FP-relation distance has an uncertainty of 23\% \citep{Said24}.  (b) Using the independent combination of distance measurements for Coma results in a reduced total of 0.061 mag.  This Comparison Total uncertainty applies when comparing a Coma-distance calibrated locally with DESI versus a Coma-distance calibrated with Planck$+$DESI.}
\label{tab:errs}
\end{deluxetable*}

\section{Discussion}\label{sec:discuss}

Although the Hubble Tension is seen as a difference in the most precise local and inferred measurements of the Hubble constant, it appears likely that the issue is more pervasive.  There appears to be a broad conflict with long-accepted distances to the nearest objects versus that inferred via the inverse, cosmological distance ladder.  Unlike most local distance ladders which rely on custom-measured distances to relatively obscure galaxies, with the DESI FP, the conflict is now seen as ``baked-in'', existing  {\it a priori} of the Hubble tension in the long history of distance measurements to Coma and without reference to any one source, method, group, or metric.  Rather than simply a conflict between two sets of measurements, it appears likely to extend to our pre-existing knowledge of local distances measured independently of their redshifts.  As the inverse distance ladder is extended ever closer to home, we may see a broadening of the set of distance indicators that can be used to test it.  We expect future DESI releases to amplify the issue by improving statistics while connecting the Hubble flow to even closer clusters, such as Fornax, Virgo, Leo I, etc.

\subsection{Additional Potential Sources of Uncertainty}

One source of uncertainty not discussed in this analysis is the radial size of the Coma cluster itself.  \cite{lokas_mamon} calculate that the radial size is 2.86 Mpc, which is $\sim3\%$ of the distance to Coma.  Following \cite{Carreres24,Peterson24b}, we check the size of large clusters like Coma in Uchuu N-body simulations and find large clusters like Coma have typical radial size of $\sim2$ Mpc, in rough agreement.  J21 measure the distance to a single galaxy, NGC 4874, which is one of the two brightest and most massive galaxies in the cluster, likely near the center of the galaxy cluster. Assuming the galaxies DESI measures the FP relation with are evenly distributed across the cluster, the difference between the mean location of the 226 FP-relation galaxies measured and the center is likely a small fraction of the 2.86-Mpc radial size. 

For the present analysis, we measure distances to 12 SNe spread across the cluster.  As shown in Fig.~\ref{fig:coma}, 6 of the 12 are likely located in the central core of the cluster.  The difference in mean brightness between the inner core and outer core is $0.075\pm0.082$ mag, a $<1\sigma$ difference (SNe near core being fainter).  Assuming that the distribution of SNe discovered is relatively uniform across the cluster, or that it roughly traces the distribution of galaxies used for FP-relation measurements, we should likewise expect this fractional difference in position along the Coma cluster to be subdominant to the other uncertainties in the Coma distance measurement.

We include both the statistical and systematic uncertainties in the FP relation from S24, and from fig.~8 of S24, we note that variants in their analysis that relate to the FP relation measurement could change the inferred distance of Coma by at most $3\%$.  S24 also extensively studies different bulk-flow models, and notes their impact is on the $\sim1\%$ level for $H_0$ or distance.   To reconcile the canonical distance to Coma and the DESI FP relation with the Planck value of $H_0$, the mean peculiar velocity would need to be $\sim -950$ km/s, much larger than the mean of $\sim-20$ km/s derived from bulk-flow models for the cluster \citep{Carr21}.

\subsection{Future Prospects}

S24 anticipates forthcoming DESI year 1 FP-relation data (Ross et al.~in preparation) to be
substantially larger (4k vs.~$\sim100$k elliptical galaxies), which will reduce both the statistical and systematic uncertainties. 
In addition, the upcoming DESI data should extend the FP relation to multiple, nearer clusters besides Coma such as Fornax, Virgo and Leo I, and given the success of the current analysis for one cluster, it is likely that various distance measurements (e.g., SNe, Cepheids, TRGB) can all be used to directly reach these clusters.  

As shown in the Appendix, from the years of 2019-2024, 18 SNe Ia were discovered and spectroscopically confirmed to be SNe Ia in Coma, a rate of several per year.  A number of the light curves for these SNe are sparse, and a commitment to multi-band, high cadence photometric follow-up for SNe in clusters like Coma would be strongly beneficial.  Additionally, in the last 5 years, according to TNS, 81 likely SNe were discovered in the area of Coma (not accounting for a redshift cut, that can be found in the ellipse of Fig.~\ref{fig:coma}).  Only 29 of the 81 received a spectroscopic identification.  A commitment to spectroscopic follow-up as well as for identification, could roughly double the number of usable SNe Ia compared to the last 5 years.  As shown in Table~\ref{tab:errs}, the uncertainty in the mean of SNe Ia in Coma is the largest uncertainty in the error budget.  With $\sim40$ SNe Ia measured, or fewer with better measurements, the uncertainty in the mean would be $\sim0.024$ mag, smaller than the uncertainty in the current calibration between SNe Ia and Cepheids and smaller than the FP-relation measurement of 226 galaxies in Coma.

\section{Conclusions}\label{sec:conclusion}

In this analysis, we calibrated 12 high-quality light curves of SNe Ia located in the Coma cluster,
yielding a standardized SN Ia brightness of $\overline{m_B^0} =15.712\pm0.041$ mag
and the highest precision distance to the Coma cluster to date.   The {\it HST} distance ladder calibration (R22) of the SNe~Ia luminosity places Coma at $34.97 \pm 0.05$ mag, or a distance of $98.5\pm2.2$ Mpc, consistent with historical measurements but with $\sim$2-3$\times$ the precision of any recent, individual measure.  The inverse distance ladder of the Hubble diagram from the DESI FP relation combined with $H_0$ as measured with Planck+$\Lambda$CDM  places Coma at a significantly larger distance of $D=111.8\pm1.8$ Mpc, $4.2\sigma$ beyond this direct measure and well beyond the consensus distance.   Alternatively, combining uncorrelated local distance measurements to Coma with the DESI measurement, we find $H_0=76.9\pm 2.0$ km/s/Mpc, $4.6\sigma$ from the Planck value.   This new route, the Coma distance and DESI Hubble diagram, yields another vantage point for the Hubble Tension, seen here from an even broader range of local distance indicators and independent of the SN Ia measuring the Hubble flow.  Based on this study of Coma or of compilations over the last several decades it is hard to see how Coma could be located as far as the cosmological expectation of 110--115 Mpc.

There are good prospects for improving upon this result in the near term.  Upcoming JWST programs have targeted Coma for intensive measurements (PI Jensen, GO 5989) in Cycle 3 (2025).  In addition,  dedicated spectroscopic and photometric follow-up of SNe in the Coma cluster can easily improve on the present result.  It is likely that within just a few years, the uncertainty in $H_0$ from a Coma-based distance ladder will not be dominated by uncertainties from measurements within Coma, but the calibration of those measurements elsewhere in this new distance ladder. 

\bibliographystyle{mn2e}
\bibliography{main}{}

\begin{acknowledgements}
\section*{Acknowledgements}
We thank the Templeton Foundation for directly supporting this research. D.S.~is supported by Department of Energy grant DE-SC0010007, the David and Lucile Packard Foundation, the Templeton Foundation, and Sloan Foundation. 
D.O.J. acknowledges support from NSF grant AST-2407632 and NASA grant 80NSSC24M0023. \end{acknowledgements}

\begin{appendix}

We include here a comprehensive list of all SNe found in our TNS and SIMBAD queries and whether they are included in the sample, and if not, the reason.  `Unavailable' means there is no light-curve photometry available.  `Too sparse' means there is a light curve, but only a small number of epochs. The rest of the notes should be self-explanatory.

Two SNe~Ia in Pantheon$+$, 1994S and 2020jhf, have redshifts right near the lower redshift limit of 0.015 ($z=0.01524, 0.015371$ respectively) and we do not include them as they may be in front of the cluster; the next lowest redshift of a SN Ia is $z=0.018$.  One supernova, 2015M, near the center of the cluster and well within the redshift range at $z=0.02316$, is classified as super-Chandrasekhar like \citep{Hicken07} from its spectrum (i.e., peculiar); this SN is much brighter than others in Coma and is thus not included in our list.

\setlength{\tabcolsep}{4pt}  % Adjusts the column spacing (optional)
\renewcommand{\arraystretch}{0.8}  % Adjusts the row spacing (less than 1 to reduce spacing)

\begin{deluxetable}{lcccc}
\tablecaption{SNe Ia Discovered in Coma, Sorted by Redshift\label{tab:supernova_data}}
\tablehead{
\colhead{Name} & \colhead{RA (deg)} & \colhead{DEC (deg)} & \colhead{$z_{\textrm hel}$} & \colhead{Note}
}
\startdata
1994S & 187.8410 & 29.1344 & 0.0152 & Very bright, in front of cluster \\
2020jhf & 198.7227 & 27.0086 & 0.0154 & Very bright, in front of cluster \\
2023czd & 195.6467 & 27.4393 & 0.0180 & In sample \\
2010ai & 194.8501 & 27.9964 & 0.0184 & In sample \\
2019bkh & 197.2389 & 28.2813 & 0.0195 & In sample \\
2021wad & 195.1308 & 28.3464 & 0.0199 & Not enough data \\
2019be & 195.0604 & 27.9568 & 0.02 & No light curve, 91bg-like \\
2020ags & 194.2706 & 29.0887 & 0.02 & In sample \\
2024ana & 194.5941 & 27.9662 & 0.0201 & In sample \\
2013ag & 192.8959 & 26.6293 & 0.0213 & In sample \\
2021dch& 196.3490 & 29.5096 & 0.0202 & In sample \\
ASASSN-16bg & 194.8546 & 27.7403 & 0.0202 & Unavailable \\
ASASSN-16np & 189.1800 & 26.1135 & 0.0208 & Unavailable \\
2024rkc & 198.3561 & 27.794 & 0.0210 & Unavailable \\
ASASSN -14bd & 193.1867 & 26.4703 & 0.0213 & Unavailable \\
2016iuc & 198.3608 & 27.8068 & 0.0214 & Unavailable \\
2024fxl & 190.1514 & 26.5059 & 0.0220 & Only one filter \\
ASASSN -15jt & 197.0380 & 27.8263 & 0.0222 & Unavailable \\
2021oat & 195.0344 & 28.1703 & 0.0225 & In sample \\
2022frn & 194.9657 & 27.9435 & 0.023 & In sample \\
2015M & 195.1346 & 27.9781 & 0.0232 & Outlier, super-Chandrasekhar \\
2020afp & 193.9218 & 27.2500 & 0.0237 & $x_1$ cut, 91bg-like \\
2001cg & 193.9203 & 27.2511 & 0.0240 & Too sparse \\
2023aakj & 195.1214 & 28.4555 & 0.0241 & In sample \\
2021lxb & 195.4944 & 28.0086 & 0.0259 & In sample \\
2023ke & 194.5760 & 29.1287 & 0.0260 & Missed peak \\
PTF 11gdh & 195.1586 & 28.0567 & 0.0262 & Unavailable \\
2023epj & 194.9738 & 26.8194 & 0.0267 & In sample \\
2006bz & 195.1808 & 27.9616 & 0.0277 & $x_1$ cut, 91bg-like \\
2009L & 194.7004 & 27.6738 & 0.0280 & Unavailable \\
2006cg & 196.2597 & 28.7400 & 0.0288 & Too sparse \\
2003an & 201.9731 & 28.5081 & 0.0327 & Unavailable \\
\enddata
\end{deluxetable}

\end{appendix}

\end{document}